\documentclass[aps,prl,twocolumn,showpacs,preprintnumbers,amsmath,amssymb,superscriptaddress]{revtex4}

\usepackage{graphicx}
\usepackage{dcolumn}
\usepackage{bm}
\newcommand{\nc}{\newcommand}
\nc{\be}{\begin{equation}}
\nc{\ee}{\end{equation}}
\nc{\bea}{\begin{eqnarray}}
\nc{\eea}{\end{eqnarray}}
\nc{\bean}{\begin{eqnarray*}}
\nc{\eean}{\end{eqnarray*}}
\nc{\mb}{\mbox}
\nc{\rnc}{\renewcommand}
\nc{\vk}{\mb{\bf k}}
\nc{\vp}{\mb{\bf p}}
\nc{\vn}{\mb{\bf n}}
\nc{\vq}{\mb{\bf q}}
\nc{\rr}{\mb{\bf r}}
\nc{\vz}{\hat {\mb{\bf z}}}
\nc{\vj}{\mb{\boldmath$j$}}
\nc{\vg}{\mb{\boldmath$g$}}
\nc{\x}{\mb{\boldmath$x$}}
\nc{\A}{\mb{\boldmath$A$}}
\nc{\va}{\mb{\boldmath$a$}}
\nc{\vs}{\mb{\boldmath$\sigma$}}
\nc{\vpi}{\mb{\boldmath$\pi$}}
\nc{\nab}{\nabla}
\nc{\X}{\sf x}

\begin{document}


\title{Quantum Hall Ferromagnetism in Graphene}
\author{Kentaro Nomura}
\affiliation{Department of Physics, University of Texas at Austin,
Austin TX 78712-1081, USA}
\author{Allan H. MacDonald}
\affiliation{Department of Physics, University of Texas at Austin,
Austin TX 78712-1081, USA}

\date{\today}

\begin{abstract}
Graphene is a two-dimensional carbon material with a honeycomb
lattice and Dirac-like low-energy excitations.  When Zeeman and spin-orbit 
interactions are neglected its Landau levels are four-fold 
degenerate, explaining the $4 e^2/h$ separation between quantized 
Hall conductivity values seen in recent experiments.  In this paper we
derive a criterion for the occurrence of interaction-driven quantum Hall
effects near intermediate integer values of $e^2/h$ due to charge gaps in
broken symmetry states. 
\end{abstract}

\maketitle

\noindent

\noindent

\begin{figure}[b]
\begin{center}
\includegraphics[width=0.4\textwidth]{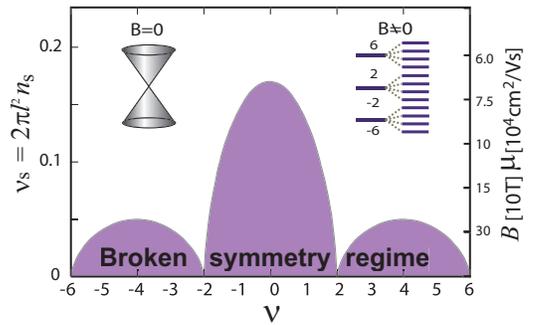}
\caption{
Phase Diagram for $SU(4)$ quantum Hall ferromagnetism in the $n=0$ and $n=1$ Landau
levels of graphene.  In our model the ordered region is bounded by a maximum value of $\nu_s$, the ratio
of the density of Coulomb scatterers to the density of a full Landau level.  $\nu_s$ is inversely
proportional to the product of the sample mobility and the external field strength and order near integer
filling factors requires the minimum values for this product indicated on the right-hand vertical axis.
}
\label{QHFPhaseDiagram}
\end{center}
\end{figure}

\noindent
{\em Introduction}---
Two-dimensional graphite (graphene) is a gapless semiconductor with a
honeycomb lattice and an unusual 
massless Dirac-fermion band structure\cite{divincenzo} that has long attracted 
theoretical attention.  The topology of its 
Bloch states leads to large momentum-space Berry phases\cite{Berry},
quantized and half-quantized Hall effects, 
and a vanishing density of states at the neutral Fermi energy which qualitatively
alters the way in which electron-electron interactions\cite{fermiliquidtheory}
influence electronic properties.  The integer quantum Hall effect in graphene 
is expected to be unusual because its Landau levels 
are widely separated and fourfold degenerate in the absence of weak 
Zeeman and spin-orbit interactions.  Interest in graphene has increased 
recently because of experimental progress\cite{graphene_exp}, including the
discovery of the integer quantum Hall effect\cite{graphene_qhe} with quantized values
of the Hall conductivity ($ \sigma_{xy}=4(n+1/2) (e^2/h)$) separated by $4 e^2/h$.  
In this Letter we address the quantum Hall effects that
should occur at intermediate integer values of filling factor 
$\nu$, giving rise to plateaus at $\sigma_{xy}=\nu (e^2/h)$, in principle for all integer values of $\nu$.
These additional plateaus are expected to arise from 
charge gaps induced by electron-electron interactions, but have not yet been observed.
They would be a new example\cite{graphene_qhf} of the enhanced 
interaction physics that occurs at integer filling factors in a strong 
magnetic field whenever $N \ge 2$ Landau levels are degenerate or nearly
degenerate.  At integer filling factors the mean-field-theory scenario in which symmetries are broken to open
gaps between quasiparticle orbitals usually applies.  The ground state is then well approximated by 
an unrestricted Hartree-Fock state\cite{jungwirth} in which an integer number $i < N$ of Landau levels 
associated with orthogonal $SU(N)$ spinors is occupied.  The phenomenon of 
interaction induced gaps and broken symmetries at integer filling factors is known
as quantum Hall ferromagnetism.  The four-fold degeneracy 
of graphene's Landau levels follows from approximate spin-degeneracy and from 
Bloch state degeneracy between two inequivalent points in the honeycomb lattice 
Brillouin zone.  The low-energy physics of graphene is well described\cite{fermiliquidtheory} in a  
four-component spinor envelope-function formalism with $SU(4)$ invariant electron-electron interactions.
Graphene is therefore a good example of $SU(4)$ quantum Hall ferromagnetism\cite{su4},
much more accurately approximating this symmetry than bilayer electron systems\cite{burkov} for example. 
The absence of additional integer Hall plateaus due to quantum Hall ferromagnetism in all 
but the most recent samples is almost certainly due to disorder. Fig.~\ref{QHFPhaseDiagram} summarizes
the estimate of the minimum sample mobility required to see quantum Hall ferromagnetism in graphene
which is explained below.

\noindent
{\em Massless Dirac-Weyl quasiparticles}---
The ${\vec k} \cdot {\vec p}$ Hamiltonian of the graphene bands is 
\begin{equation}
H_0= v (p_x + eA_x)\tau_z\sigma_x+v (p_y + eA_y)\sigma_y
\label{dh1c}
\end{equation}
where $\tau_z = \pm$ labels the two-degenerate ($K$ and $K'$)  
valleys, $\sigma_{\alpha}$ are Pauli matrices that act in the space of the two-atoms 
per unit cell, and ${\vec A}({\vec r})$ is the vector potential.
In the zero field case, the Hamiltonian (\ref{dh1c}) has linear dispersion
$E=\pm v\hbar k$ for both spin states and for both $K$ and $K'$ valleys.
In a magnetic field the spectrum of $H_0$ consists of four-fold degenerate (including spin) 
Landau level branches with $E_n =  \pm \hbar v \sqrt{2n}/\ell$ as
indicated in the inset in Fig.1.
For $n = 0$ eigenfunctions of different valleys are localized on different honeycomb sublattices
while for $n \ne 0$, they are symmetric or antisymmetric combinations of two-dimensional-electron-gas
Landau level $n$ states on one sublattice and level $n+1$ states on the other sublattice.  The spin
degeneracy is lifted by weak Zeeman coupling that we neglect for the moment.

\noindent
{\em Stoner Criterion}--- The simplest approximation\cite{disorderplusexchange} for interacting electrons in a strong 
magnetic field is one in which interactions are treated in the Hartree-Fock approximation
and disorder in the self-consistent Born approximation\cite{disorderplusexchange}(SCBA). 
In the strong field limit the total 
energy in this approximation is given by 
\begin{equation} 
\frac{E}{N_{\phi}} = \sum_{\sigma=1}^{4} \; [\, \int_{-\Gamma}^{\mu_{\sigma}} \, dE \, E\, A(E)  - \frac{X}{2} \nu_{\sigma}^2 \, ] 
\label{totaleng}
\end{equation} 
where $N_{\phi}$ is the orbital Landau level degeneracy, 
$\mu_{\sigma}$ is the Fermi level for spinor-component $\sigma$, 
\begin{equation}
\label{xint} 
X = \int \frac{d^2 \vec q}{(2\pi)^2} \; V_{int}({\vec q}) \; \exp(-q^2\ell^2/2) \; F^2(\vec{q})
\end{equation}
is the exchange integral, $\ell = (\hbar c/eB)^{1/2}$ is the magnetic length, 
and $F(\vec{q})$ is a form factor we will discuss later.
In the SCBA the Landau level spectral function has the form 
\begin{equation} 
\label{dos}
A(E) = \frac{2}{\pi \Gamma} \; [1 - (E/\Gamma)^2\,]^{1/2}, 
\end{equation} 
where
$\Gamma$ is the Landau level width\cite{disorderplusexchange}:
\begin{equation}
\label{disint}
\frac{\Gamma^2}{4} = n_{s} \int \frac{d^2 \vec q}{(2\pi)^2} \;
|U_{dis}({\vec q})|^2 \; \exp(-q^2\ell^2/2) \; F^2(\vec{q})  
\end{equation} 
which is estimated below.
In Eqs.(\ref{xint}) and (\ref{disint}) $V_{int}({\vec q})$ and $U_{dis}({\vec q})$ are 
the Fourier transforms of the electron-electron interaction and the disorder potential
and $n_{s}$ is the density of disorder scatterers. 
In the case of graphene the spinor index
$\sigma$ runs over four possible values.  These expressions assume perfect $SU(4)$ invariance
of the disorder-scattering and electron-electron interactions.  While this is certainly an 
approximation, we believe it to be an excellent one.  They also assume that
spatial invariance is recovered after disorder averaging, so that the electron density matrix is 
diagonal in its orbital labels and the energy simply proportional to the number of orbitals in a 
Landau level $N_{\phi}$.  In Eq.(~\ref{totaleng}) the $\nu_{\sigma}$ values are the eigenvalues
of the density matrix in spinor space, 
\begin{equation}
\nu_{\sigma} = \int_{-\Gamma}^{\mu_{\sigma}} dE \ A(E),
\end{equation} 
which are invariant under unitary transformations of the four-dimensional spinor-space, consistent
with $SU(4)$ symmetry.

In the normal state the four (nearly) degenerate Landau levels
are equally occupied.  To look for broken symmetry states we write
$\nu_{\sigma} = \frac{\nu_{T}}{4} + \delta \nu_{\sigma}$ 
where $\nu_{T} = N/N_{\phi}$ is the total filling factor in the four-fold degenerate 
Landau level of interest.  Expanding to second order in 
$\delta \nu_{\sigma}$ and using that $\sum_{\sigma} \delta \nu_{\sigma} = 0$ we find that
\begin{eqnarray} 
\frac{E}{N_{\phi}} &=& \sum_{\sigma} \; \big[ \, \int_{-\Gamma}^{\mu_0} \, dE \, E\, A(E) - \frac{X}{8} \nu_T^2 \, \big] \nonumber \\
&+& \sum_{\sigma} \; \frac{\delta \nu_{\sigma}^2}{2} \;  \big[ \frac{1}{A(\mu_0)} - X \big] + \ldots
\label{changenergy}
\end{eqnarray} 
where $\mu_0$ is the normal state Fermi level.
The normal state is unstable when the second term in square brackets is negative, in other words when
$X A(\mu_0) > 1$.  This criterion for quantum Hall ferromagnetism is closely analogous to the 
Stoner\cite{Stoner} criterion for ferromagnetism in metals, and has been successfully applied\cite{maude} to 
understand the appearance of spin-splittings at odd integer filling factors in a semiconductor
two-dimensional electron gas.  In the case of quantum Hall ferromagnetism (QHF) the exchange energy competes with
disorder energy rather than with band energy.  We can apply the QHF Stoner criterion to graphene by 
relating the disorder potential to the zero-field mobility of graphene, a quantity that is conveniently available from
experiment.

\noindent
{\em Zero-Field Mobility and Coulomb Scattering}---We start from the Boltzmann transport theory expression for the 
conductivity, applied to the four-fold degenerate Bloch bands of graphene:
\begin{equation} 
\label{boltzmann}
\sigma_{B=0}^{} = \frac{e^2 \tau v^2 {\cal D}(E_F)}{2} = \frac{e^2}{h} \; \frac{2 E_F \tau}{\hbar} 
\end{equation} 
where $\tau$ is the scattering rate,
\begin{equation}
\label{scatteringrate}  
\tau^{-1} = \frac{n_s k_F}{2 \pi \hbar^2 v} \int_{0}^{2\pi} d \theta\;  |U_{dis}(q)|^2  \;
(1-\cos\theta) \; \frac{(1 + \cos\theta)}{2} 
\end{equation} 
$\theta$ is the scattering angle, $k_F$ is the Fermi wavevector and $q=2k_F \sin(\theta/2)$ is the scattering wavevector on the 
circular two-dimensional Fermi surface.  The last $\theta$-dependent factor in 
Eq.(~\ref{scatteringrate}) is non-standard and is due to the wavevector dependence of 
the relative phase of graphene Bloch band wavefunctions on the two sites within its 
honeycomb lattice unit cell.  The factor of $k_F$ in Eq.(~\ref{scatteringrate}) 
reflects the density dependence of the density-of-final states for elastic scattering of 
Fermi surface quasiparticles.  For short-range scatterers the integral in Eq.(~\ref{scatteringrate})
remains finite as density $\propto k_F^2$ vanishes.  Since ${\cal D}(E_F)$ is 
proportional to $k_F$ for two-dimensional Dirac bands, Eq.(~\ref{scatteringrate}) implies a conductivity 
that is independent of $k_F$ and therefore independent of carrier density.  Indeed theoretical studies of the 
conductivity of graphene\cite{condtheory} predict that the conductivity has a weak density dependence, remaining 
finite as $k_F \to 0$.
Experiment, on the other hand, finds that the mobility $\mu = - \sigma / ne$ 
in graphene is nearly constant except at very low-densities and that it has values 
$\sim 10^4 {\rm cm}^2 {\rm V}^{-1} {\rm s}^{-1}$ in samples that are sufficiently high quality to exhibit
the integer quantum Hall effect.  Evidently quasiparticle scattering amplitudes are enhanced at lower densities 
in such a way as to convert the $k_F^{+1}$ dependence of the scattering rate in Eq.(~\ref{scatteringrate}) to a 
$k_F^{-1}$ dependence.  One plausible explanation for this behavior is that Dirac band quasiparticle scattering
is dominated by Coulomb scattering from charged defects near the graphene plane.  For two-dimensional 
graphene $U_{dis}(q) = V_{C}(q) = 2 \pi e^2/q$.  Inserting this expression in Eq.(~\ref{scatteringrate}) 
we obtain that
\begin{equation}
\frac{E_F \tau}{\hbar} = \frac{n}{n_s} \; \frac{4}{ \pi g^2} 
\label{bareCoulombscattering}
\end{equation}  
where
$
g = \frac{e^2}{\hbar v} \sim 3
$ 
is the effective {\em fine structure constant} used to characterize the ratio of Coulomb interaction 
and band energy scales in graphene.  In Eq.(~\ref{bareCoulombscattering}) 
$n_{S}$ should be thought of as the density of Coulomb scatterers that
are located in the substrate within a Fermi wavelength of the graphene plane.  The influence of 
more remote scatterers is suppressed by the factor $\exp(-qd)$ that appears
in the two-dimensional Fourier transform of the Coulomb interaction.  
Inserting Eq.(~\ref{bareCoulombscattering}) in Eq.(~\ref{boltzmann}) we find that 
mobility
\begin{equation} 
\mu \sim \frac{ 170 \; {\rm cm}^2 {\rm V}^{-1} {\rm s}^{-1}}{n_s \; [10^{11} {\rm cm}^{-2}] }. 
\end{equation} 

In systems with Coulomb electron-electron or electron-impurity interactions screening 
normally plays an essential role, changing long-range interactions into short-range ones.
In a static approximation, the screened disorder potential in graphene is 
\begin{equation} 
U_{sc} (q) = \frac{ 2 \pi e^2} { q + 2 \pi e^2 \Pi (q)} 
\label{screening}
\end{equation} 
where $\Pi(q)$ is the polarization function of the graphene Dirac bands.  Screening 
does not change the density dependence of the conductivity in graphene because 
$\Pi(q)$ also scales like $k_F$.  The influence of screening on the mobility 
can be estimated by making a Tomas-Fermi approximation, replacing $\Pi(q)$ 
by $\Pi(q=0) = {\cal D}(E_F)$.  When the coupling constant 
$g$ is much larger than 1 $U_{sc}(q) \simeq (\hbar v \pi)/(2 k_F)$ and 
\begin{equation}
\frac{E_F \, \tau}{\hbar} \simeq \frac{n}{n_s} \; \frac{64}{\pi}
\label{screened}
\end{equation} 
yielding a value for the mobility that is $ 16 g^2$ times larger than the unscreened 
value.  
We note that $g$ cancels in Eq.(~\ref{screened}), which is fortunate because its effective value can 
be influenced by non-universal substrate dielectric screening. 
Corrections to Eq.(~\ref{screened}) becomes impotant for $g<1$.
We can use these expressions
to extract a value for the density of scatterers $n_s$ from measured mobilities.
This procedure might retain partial validity, depending on the details, even if the limiting 
scatterers are not Coulombic.  Similar density dependence could in principle arise from 
a partially accidental combination of disorder sources that gives rise to a similar
increase in transition rates at small wavevectors, or from strong short-ranged 
scattering that approaches the unitary limit.  Other potential disorder sources include 
random crumpling of the graphene sheet and coordination defects in the graphene 
sheet that give rise to long range strain fields.  The procedure 
we now use to translate between zero-field mobilities and strong-field Landau level
widths will retain its validity in some, but not all, plausible scenarios.     
In particular the values of $\mu B$ at the stoner phase boundary are
likely to be similar for Coulomb and topological defect scatterng\cite{topological}.

\noindent
{\em Self-Consistent Screening in a Magnetic Field}--- We are now in a position to estimate the 
Landau level width and apply the Stoner criterion.  For Coulomb scattering 
\begin{equation}
\label{disintCoulomb}
\frac{\Gamma^2}{4} = n_{s} \int \frac{d^2 \vec q}{(2\pi)^2} \;
\left|\frac{2\pi e^2}{q+2\pi e^2\Pi({\vec q})}\right|^2 \; \exp(-q^2\ell^2/2) \;
F^2(\vec{q}).  
\end{equation}  
Notice that $\Gamma^2$ diverges if we neglect screening.  We now need to specify the form factor 
$F(q)$.   
Taking the Coulomb interaction to be diagonal in honeycomb lattice site index it 
follows\cite{disorderplusexchange} that the form factor $F(q) \equiv 1$ for $n=0$ and 
that 
\begin{equation} 
F(q) = \frac{1}{2} \; \big[ L_{|n|}(q^2 \ell^2/2) + L_{|n|-1}(q^2 \ell^2/2) \big]
\label{formfactor}
\end{equation} 
for $n \ne 0$.  If the magnetic field is strong enough
to neglect coupling between different Landau levels the normal state polarization function
$\Pi(q)|$ is given approximately by 
\begin{equation}
\Pi(q) \approx \frac{ 4 \exp(-q^2 \ell^2/2) }{2 \pi \ell^2} \; A(\mu_0). 
\label{strongfieldPolarization} 
\end{equation} 
The factor of $4$ in Eq.(~\ref{strongfieldPolarization}) is the graphene 
Landau level degeneracy and the factor $\exp(-q^2 \ell^2/2)$ accounts for the orbital character
of Landau level wavefunctions.  Since $A(\mu_0)$ is 
proportional to $\Gamma^{-1}$, Eq.(~\ref{disintCoulomb}) must be solved self-consistently\cite{scscreening}
giving rise to the following implicit equation for $\tilde{\Gamma}  \equiv \Gamma/(e^2/\ell)$:
\begin{equation} 
1= 4\nu_s \int_{0}^{\infty} dx \frac{ \; F^2\exp(-x^2/2) \;}{\big[\; \tilde{\Gamma} x  +  F^2 4 \tilde{A}_0 \, \exp(-x^2/2
) \; \big]^2}. 
\label{selfconsistentscreening}
\end{equation}  
In Eq.(~\ref{selfconsistentscreening}) $\tilde{A}_0 \equiv \Gamma A(\mu_0)$
and $\nu_s = 2 \pi \ell^2 n_s$ is the 
`filling factor' of scatterers.  Note that since the right hand side of Eq.(~\ref{selfconsistentscreening}) 
is a monotonically increasing function of $\nu_s$ and a monotonically decreasing function of $\tilde{\Gamma}$,
$\tilde{\Gamma}$ must increase monotonically with $\nu_s$. 

\noindent
{\em Graphene QHF Phase Boundary}--- The Stoner criterion can be written in terms of $\tilde{\Gamma}$, 
$\tilde{A}_0$, and the dimensionless exchange integral 
\begin{equation}
\tilde{X} \equiv \frac{X}{e^2/\ell} = \int_{0}^{\infty} dx
 \frac{\tilde{\Gamma} x F^2\exp(-x^2/2) }  {\tilde{\Gamma} x  +  4 
\tilde{A}_0 F^2 \, \exp(-x^2/2) } 
\end{equation} 
In the absence of screening (large $\tilde{\Gamma}$), $\tilde{X}=\sqrt{\pi/2}$ for 
$n=0$ and $\tilde{X} = (11/16) \; \sqrt{\pi/2}$ for $n=1$.  
The Stoner criterion,
\begin{equation} 
\tilde{X} \tilde{A}_0 / \tilde{\Gamma} > 1,
\label{StonerDimensionless}
\end{equation}
is satisfied for $\nu_s > \nu_s^*$.
Since $\nu_s \propto n_s/B \propto 1/\mu B$, our Stoner criterion specifies the minimum
values for the product of field and mobility illustrated in Fig.~\ref{QHFPhaseDiagram}:
\begin{equation} 
B \; [10 Tesla] \; \mu \; [10^4 {\rm cm}^2 {\rm V}^{-1} {\rm s}^{-1} ] \;\;  \gtrsim 1/\nu_s^*(\nu_T).
\end{equation} 
The origin of the weaker tendency to ordered states in the four-fold degenerate $n=1$ Landau level 
is the difference in form factor $F(q)$.  

\noindent
{\em Discussion}--- Although the Stoner criterion can be applied at all
filling factors and provides a reasonable assessment of the crossover between
interaction dominated and disorder dominated physics, we caution that the 
simple quantum Hall ferromagnet states can occur only at integer values of 
the total filling factor.  We expect the emergence of interaction-driven
gaps at intermediate integer filling factors to be the first signal that sample
quality is adequate to see interaction dominated physics.  Judging by the relative
size of charge gaps at integer and fractional filling factors, we expect that the 
first fractional filling factors will require mobilities approximately five times 
larger than those required to realize quantum Hall ferromagnetism; the $SU(4)$ nature
of these Landau levels will open up a new frontier for the fractional quantum Hall 
effect that is likely to yield some surprises.  We have so far neglected the 
Zeeman energy because it is much weaker than the competing disorder and interaction
energy scales and will have little influence on whether or not quantum Hall ferromagnetism
occurs.  (The Zeeman energy is $\sim 1 {\rm meV}$ at $\sim 10 {\rm Tesla}$ compared to 
a $n=0$ Landau level interaction energy scale $\sim 100 {\rm meV}$ depending on the 
degree of substrate dielectric screening.)  When quantum Hall ferromagnetism does occur,
however, the Zeeman energy will play a larger role.
For $\nu=\pm 1$, in particular, the Zeeman energy will select ordered states that are
spin-polarized, and break symmetry in the $SU(2)$ valley space.  The interaction terms in 
graphene should be weakly dependent on valley index, because interactions on the same 
graphene sublattice should be more strongly repulsive than interactions between sublattices at 
short distances, reducing the broken symmetry to $U(1)$.
For this reason, we anticipate that the $\nu= \pm 1$ quantum Hall
ferromagnet in graphene should have a Kosterlitz-Thouless phase transition at a low temperature.
Finally we compare our result with the recent experiment\cite{graphene_qhf} that 
has reported quantum Hall ferromagnetism in graphene.
The mobility of the sample used in \cite{graphene_qhf} is
$\mu=5\times10^{4}[cm^2/Vs]$. Fig.[~\ref{QHFPhaseDiagram}] indicates that
for this mobility and $\nu=\pm 1$ the symmetry breaks at 17T in agreement with experiment\cite{graphene_qhf}.  The appearance of quantum Hall plateaus observed at 
$\nu=\pm 4$, in the middle of the four-fold degenerate $n = \pm 1$ Landau levels
at around $30 {\rm Tesla}$ is also in reasonable agreement with
Fig.[~\ref{QHFPhaseDiagram}] giving the critical field 40T.
The influence of dielectric screening on our phase diagram, which we expect to be rather weak,
and of screening due to virtual inter-Landau-level transitions, which we expect to be important 
at larger $n$, will be discussed in subsequent work.

\noindent
{\em Acknowledgments}--The authors acknowledge helpful interactions with 
Luis Brey, Andre Geim, Jason Hill, Zhigang Jiang, Philip Kim, Antonio
Neto, Nuno Peres, Nikolai Sinitsyn, and Jairo Sinova. 
 This work has been supported by 
the Welch Foundation and by the Department of Energy under grant
DE-FG03-02ER45958.

\end{document}